\documentstyle[aps,prl,floats,epsf,twocolumn]{revtex}

\begin{document}

\draft
\wideabs{
\title{Insulating and Fractional Quantum Hall States in the N=1 Landau Level}

\author{J.~P. Eisenstein$^1$, K.~B. Cooper$^1$, 
L.~N. Pfeiffer$^2$, and K. W. West$^2$}
\address{$^1$California Institute of Technology, Pasadena CA 91125 \\
         $^2$Bell Laboratories, Lucent Technologies, Murray Hill, NJ 07974}

\maketitle

\begin{abstract}
The observation of new insulating phases of two-dimensional electrons in the 
first excited Landau level is reported.  These states, which are manifested as re-entrant integer quantized Hall effects, exist alongside well-developed even-denominator fractional quantized Hall states at $\nu=7/2$ and 5/2 and new odd-denominator states at $\nu=3+\frac{1}{5}$ and $3+\frac{4}{5}$.
\end{abstract}

\pacs{73.43.Qt, 73.20.Qt, 73.63.Hs}
}

At zero magnetic field Coulomb interactions in two-dimensional electron systems 
(2DES) are less important than kinetic effects at all but the lowest densities.  
A large magnetic field $B$ applied perpendicular to the 2D plane fundamentally 
changes this situation by resolving the kinetic energy spectrum into a ladder of 
discrete, yet highly degenerate, energy levels.  In the absence of disorder 
these Landau levels (LL) have zero width, making electron-electron interactions 
impossible to ignore.  At fields high enough that all electrons reside in the 
lowest ($N=0$) Landau level, interactions produce a plethora of incompressible 
quantized Hall phases at odd-denominator fractional fillings of the level, 
intriguing metallic phases at certain even-denominator fillings, and insulating states at very small fillings which presumably reflect pinned Wigner crystals\cite{perspectives}. 

At lower magnetic fields, where the excited LLs are populated, a very different 
situation prevails.  Recent experiments have shown that when three or more LLs 
are occupied (i.e. $N\ge2$) remarkable anisotropic collective states develop 
near half filling and curious isotropic insulating phases appear in the flanks 
of the LL\cite{lilly1,du}.  These new states, which are quite distinct from the 
fractional quantized Hall fluids found in the $N=0$ LL, are observed only in very high quality samples and at very low temperatures. Although their precise nature 
remains unclear, there is widespread belief that they are ``stripe" and 
``bubble" charge density wave (CDW) or liquid crystalline phases closely related 
to those originally predicted by Hartree-Fock theory\cite{KFS,MC}. 

Between these two regimes lies the $N=1$ first excited LL. The discovery of an even-denominator fractional quantized Hall effect (FQHE) at Landau level filling fraction $\nu=5/2$ gave the first indication that physics in the $N=1$ LL was different from that in the $N=0$ level\cite{willett}. Since then only a small number of additional FQHE states (at $\nu=7/3$, 8/3 and 7/2) have been observed. Most recently it was reported that the 5/2 and 7/2 FQHE 
states are replaced by strongly anisotropic phases when a substantial magnetic 
field component $B_{||}$ parallel to the 2D plane is added to the existing 
perpendicular field\cite{pan,lilly2}.  These anisotropic phases appear to be 
similar to those found at $B_{||}=0$ in the $N\ge2$ LLs. Indeed, Rezayi and Haldane have argued that the parallel field drives a transition from a paired quantized Hall state to a stripe phase similar to those found in the $N\ge2$ LLs\cite{rezayi}.   Thus, the $N=1$ LL exhibits phenomena characteristic of both the $N=0$ FQHE and the $N\ge2$ CDW/liquid crystal regimes as well as its own unique features.  In this paper we report the observation of several new phases in the $N=1$ LL.  Most importantly, {\it eight} isotropic insulating states exhibiting integer Hall quantization have been found interposed between various even- and odd-denominator fractional quantized Hall states.  These insulating states develop only below $T\sim 50 \ {\rm mK}$.

\begin{figure}[h]
\begin{center}
\epsfxsize=3.3in
\epsfclipon
\epsffile[163 222 419 560]{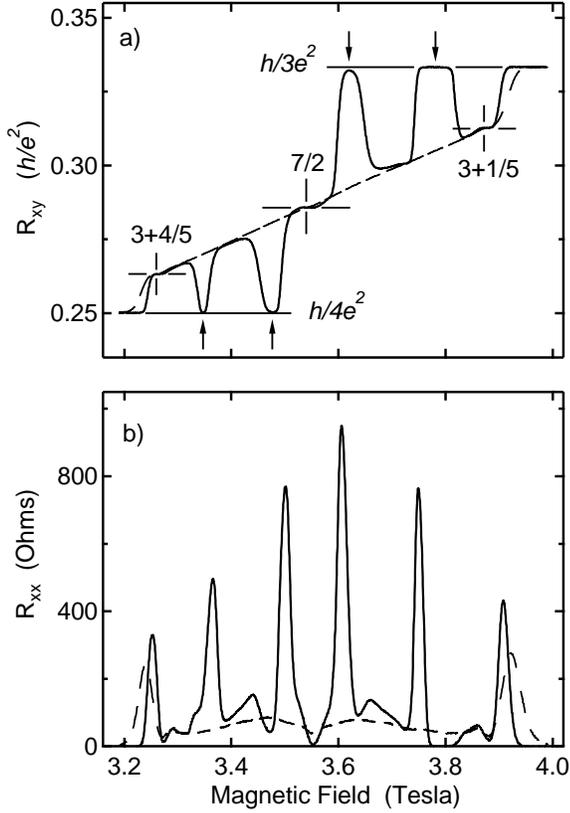}
\end{center}
\caption[figure 1]{a) Hall resistance in the upper spin branch of the $N=1$ Landau level.  Solid curve, $T\approx 15 \ {\rm mK}$; dashed curve $T=50 \ {\rm mK}$. Cross hairs indicate locations of FQHE states at $\nu=7/2$, $3+\frac{1}{5}$ and $3+\frac{4}{5}$. Arrows denote locations of new insulating states. b) Longitudinal resistance data at same magnetic fields and temperatures.
}
\end{figure}
Figure 1 shows magneto-transport data from a 2DES sample consisting of a single 
30 nm-wide modulation-doped GaAs quantum well.  After brief low temperature 
illumination with a red light emitting diode, the 2D electron gas in this sample 
has a density of $N_s=3.0\times 10^{11} {\rm cm^{-2}}$ and a mobility of $\mu \approx 3.1 \times 10^7 {\rm cm^2/Vs}$.  The data in Fig. 1, which span the magnetic field range from $B=3.2$ to $4.0{\rm T}$ corresponding to Landau level filling factors between $\nu \equiv hN_s/eB \approx 4$ and 3, reflect properties of the 2D system in the upper spin branch of the $N=1$ Landau level.  Both Hall 
($R_{xy}$) and longitudinal ($R_{xx}$) data, taken at our dilution 
refrigerator's base temperature of about $T \approx 15 \ {\rm mK}$ and at $T=50 \ {\rm mK}$, are shown.

The Hall resistance data at $T \approx 15 \ {\rm mK}$ shown in Fig. 1a contain 
several notable features.  At the upper and lower magnetic field extremes well-defined plateaus at $R_{xy} =h/3e^2$ and $h/4e^2$, respectively, are evident.  
These plateaus correspond to the conventional integer quantized Hall effects at 
Landau level filling factors $\nu=3$ and 4. Near the center of the field range, 
at $B=3.54 \ {\rm T}$, a flat Hall plateau is apparent.  The Hall resistance on this plateau equals $2h/7e^2$ to within 0.015\%, thus verifying the existence of an even-denominator FQHE at $\nu=7/2$.  Accurately quantized Hall plateaus corresponding to FQHE states at $\nu=3+\frac{1}{5}$ and $3+\frac{4}{5}$ are also present.  Perhaps surprisingly, the data offer only weak indications of FQHE states at $\nu=3+\frac{1}{3}$ and $3+\frac{2}{3}$.

The most remarkable features in Fig. 1a are the four isolated regions, denoted 
by arrows, in which the Hall resistance approaches either $h/3e^2$ or $h/4e^2$.  
The non-monotonic variation of $R_{xy}$ in these regions readily distinguishes 
them from the conventional $\nu =3$ and $\nu =4$ integer quantized Hall states. 
Of the four regions, the one near $B=3.78 \ {\rm T}$ is the best developed; the data show $R_{xy}=h/3e^2$ to within 0.02\%.  The regions near $B=3.62$ and 3.48 T, which straddle the $\nu =7/2$ FQHE state, are clearly beginning to develop plateaus at $R_{xy}=h/3e^2$ and $h/4e^2$, respectively.  The lowest field feature, near $B=3.35 \ {\rm T}$, is the weakest of the four, but it seems quite likely that $R_{xy}$ is approaching $h/4e^2$.  These four features are 
very  temperature sensitive; the dashed trace in Fig. 1a shows that they are 
completely absent at $T=50 \ {\rm mK}$. 

\begin{figure}[h]
\begin{center}
\epsfxsize=3.3in
\epsfclipon
\epsffile{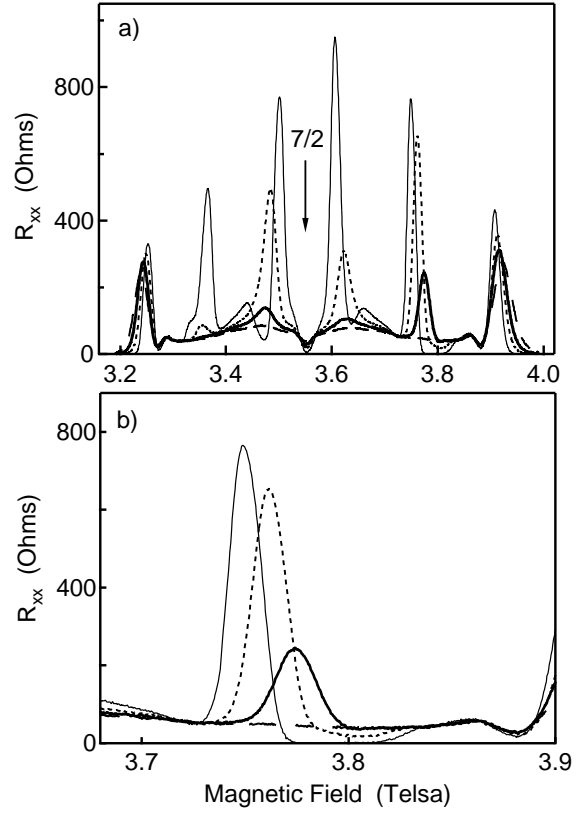}
\end{center}
\caption[figure 2]{a) Temperature dependence of the longitudinal resistance in the upper spin branch of the $N=1$ Landau level $4>\nu>2$.  Light solid curve: 15 mK; dotted curve: 30 mK; heavy solid curve: 40 mK; and dashed curve: 50 mK. 
b) Expanded view of $R_{xx}$ associated with the re-entrant integer Hall quantized effect near $B=3.78 \ {\rm T}$.
}
\end{figure}

Figure 1b shows longitudinal resistance data from the same field range as the 
Hall data in Fig. \ 1a.  At $B=3.54 \ {\rm T}$ a deep minimum in $R_{xx}$ associated 
with the FQHE state at $\nu=7/2$ is evident.  The observed temperature 
dependence of $R_{xx}$ at this magnetic field is consistent with thermal 
activation, $R_{xx}\!\propto e^{- \Delta /2T}$; we estimate $\Delta_{7/2}\! \approx\! 70 \ {\rm mK}$.  The Hall plateaus associated with the $\nu=3+\frac{1}{5}$ and $3+\frac{4}{5}$ FQHE states are also accompanied by well-defined minima in $R_{xx}$. 

The longitudinal resistance in the vicinity of each of the four isolated regions 
where the Hall resistance approaches either $h/3e^2$ or $h/4e^2$ exhibits 
somewhat intricate magnetic field and temperature dependences.  These are 
illustrated in Fig. 2a where $R_{xx}$ data taken at $T=50$, 40, 30, and 
approximately 15 mK, are displayed.  Concentrating first on the narrow field 
range near $B=3.78 \ {\rm T}$ which is expanded in Fig. 2b, the data show 
that a peak in $R_{xx}$ develops as the temperature is reduced from $T=50 \ {\rm 
mK}$ to 40 mK. By 30 mK this peak has grown substantially, shifted to slightly 
lower magnetic field, and has been joined by a weak relative minimum on its high 
field side.  By approximately 15 mK the peak has grown and shifted slightly more 
but the minimum has evolved into a significant region of near zero longitudinal 
resistance.  This zero in $R_{xx}$ coincides closely with the strong plateau in 
the Hall resistance at $R_{xy}=h/3e^2$ noted previously.

The behavior of $R_{xx}$ near the three other regions of re-entrant Hall 
quantization is qualitatively similar, if less well developed.  For the two 
regions straddling the $\nu=7/2$ FQHE state, peaks in $R_{xx}$ develop first upon cooling but these are eventually followed by the formation of adjacent minima.  The depth of these minima correlates well with the strength of the developing plateaus in $R_{xy}$. For the weakest feature, near $B=3.35 \ {\rm T}$, a clear peak develops and, at the lowest temperature, a shoulder on its low field side suggests an incipient minimum. 

The four regions of re-entrant integer Hall quantization are centered near Landau level filling factors $\nu \approx 3.28$, 3.42, 3.56, and 3.70.  The complex temperature dependence of $R_{xx}$ and integer quantization of $R_{xy}$ around these filling factors suggests that they are not quantum numbers in the sense of the FQHE.  Indeed, the minima in $R_{xx}$ described above are associated with the onset of insulating behavior, not the formation of new FQHE states.

Figure 3 demonstrates that similar re-entrant integer quantized Hall effect 
(RIQHE) states also appear in the lower spin branch of the $N=1$ LL, between 
filling factors $\nu =2$ and 3.  In these data both of the re-entrant Hall features above half-filling exhibit strong plateaus accurately quantized to $h/2e^2$.  Anomalous minima and maxima in $R_{xy}$ in this filling factor range have been reported before\cite{pan2}, but the present data are, to our knowledge, the first to demonstrate re-entrant integer Hall quantization.  Pan, {\it et al.}\cite{pan2} also observed large changes in the anomalous $R_{xy}$ features when the direction of the perpendicular magnetic field was reversed. This was attributed to mixing of the longitudinal and Hall resistances in the sample.  We emphasize that in our sample field reversal has a negligible effect.
Finally, a strong $\nu=5/2$ fractional quantized Hall effect is also apparent in Fig. 3.  The temperature dependence of $R_{xx}$ at $\nu=5/2$ suggests an energy gap $\Delta_{5/2} \approx 310 \ {\rm mK}$.   

The transport data described above were obtained with the magnetic field 
perpendicular to the 2D plane.  Examination of the longitudinal and Hall 
resistances for sensitivity to the direction of current flow within the 2D plane 
shows no evidence, in the $N=1$ LL, of the large systematic anisotropy which is
readily apparent in the $N \ge 2$ levels.  On the other hand, it is known
that the imposition of a magnetic field component $B_{||}$ in the 2D 
plane destabilizes the $\nu=5/2$ and 7/2 FQHE states\cite{eisenstein} and 
renders the longitudinal resistance of the sample strongly 
anisotropic\cite{pan,lilly2,hardaxis}.  By tilting the present sample relative 
to the total applied magnetic field we have corroborated these earlier findings 
and have found that the new re-entrant integer quantized Hall features reported 
here are gradually suppressed as the in-plane field is increased. 

\begin{figure}[h]
\begin{center}
\epsfxsize=3.3in
\epsfclipon
\epsffile{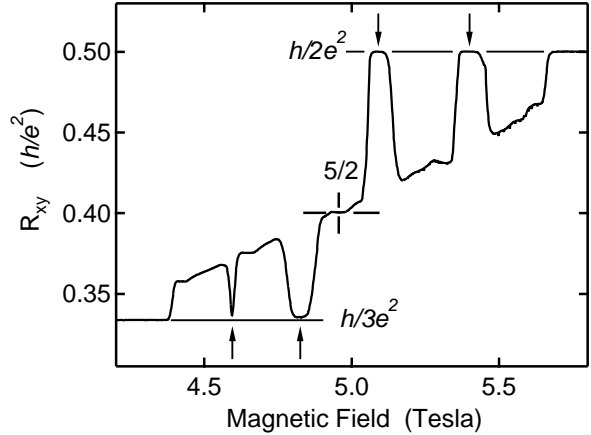}
\end{center}
\caption[figure 3]{Hall resistance data at in the lower spin branch of the $N=1$ Landau level, $3>\nu>2$, at $T\approx 15 \ {\rm mK}$.  New insulating states indicated by arrows, cross-hairs locate the $\nu=5/2$ FQHE state.
}
\end{figure}

The discovery of re-entrant integer quantized Hall states in the $N=1$ Landau 
level suggests the existence of collective insulating states.  If 
the electrons in the partially filled uppermost LL are insulating, the filled 
LLs beneath the Fermi level ensure integer quantization of the Hall resistance and 
zero longitudinal resistance.  The fact that these new insulators are 
{\it re-entrant}, i.e.\ are separated from one another and the conventional 
integer QHE states by magnetic field ranges exhibiting fractional quantized Hall states or simply finite $R_{xx}$ and non-quantized $R_{xy}$, is strong evidence that the origin of the insulating behavior is collective.  Even if the conventional integer QHE states at $\nu=3$ and $\nu=4$ are dominated by disorder-induced single-particle localization, the presence of conducting regions between them and the re-entrant insulating states necessarily implies that the quasiparticles in the system first delocalize upon moving away from integer filling. It seems highly unlikely that the same single-particle mechanism could then become operative again at the much higher quasiparticle densities present in the re-entrant insulating states. Furthermore, the existence of the FQHE states at $\nu=3+\frac{1}{5}$ and $\nu=3+\frac{4}{5}$ offers compelling evidence that electron-electron interactions are already dominant even before the first re-entrant insulator has formed. 

Of the four re-entrant IQHE states observed in the upper spin branch of the $N=1$ LL (i.e. $4 > \nu > 3$), only the one near $B=3.78 \ {\rm T}$ actually exhibits vanishing $R_{xx}$ at $T>15 \ {\rm mK}$.  As suggested above, we believe that the non-zero $R_{xx}$ of the other three states merely reflects their weaker development.  
In each of the four cases, a peak in $R_{xx}$ develops first upon cooling and is 
followed by the formation of a minimum at lower temperatures.  We speculate that  
this behavior reflects the shifting balance of current between the edge channels 
of the filled lower LL levels and the ``valence" electrons in the 
uppermost LL.  If the valence electrons are highly conducting, say at high 
temperature, then $R_{xx}$ will obviously be small. On the other hand, if the 
valence electrons are insulating, the resistance $R_{xx}$ will be essentially 
zero owing to the edge channels of the LLs beneath the Fermi level.  A peak in $R_{xx}$ is therefore expected at some intermediate value of the longitudinal conductivity of the valence electrons.  

The re-entrant insulating states reported here are reminiscent of the re-entrant 
integer QHE states recently found in the $N\ge2$ LLs\cite{lilly1,du}. Two such insulating states, one near $1/4$-filling and the other near $3/4$-filling, are observed within each spin sublevel.  There is evidence, both experimental\cite{cooper} and theoretical\cite{haldane,yoshioka} that these states are related to the ``bubble" charge density waves predicted by Koulakov, Fogler, and Shklovskii\cite{KFS} and by Moessner and Chalker\cite{MC} on the basis of Hartree-Fock theory.  The bubble phases, which consist of triangular arrays of multi-electron clusters, are interposed between unidirectional stripe phases around half filling and Wigner crystals deep in the flanks of the level. Pinning of the bubble arrays presumably produces the observed insulating behavior.  We note, however, that the theoretical work on bubble phases has so far been aimed at the $N\ge2$ LLs, not the $N=1$ level.

Re-entrant insulating states have also been observed in the $N=0$ lowest LL\cite{shayegan}.  For example, the Laughlin liquid FQHE state at $\nu=1/5$ is flanked by insulating behavior above and in a narrow window below it in magnetic field\cite{insulate}.  A plausible interpretation\cite{jiang} of this is that the insulating behavior results from the pinning of magnetic field-induced Wigner crystals, but re-entrance occurs because the downward cusp in the energy of the Laughlin liquid makes it the ground state in a narrow range of filling factor around $\nu=1/5$.  For two-dimensional {\it hole} systems, a similar re-entrance is observed\cite{manoharan} around the $\nu=1/3$ FQHE state.   The shift of the re-entrance to higher filling factors in 2D hole systems has been attributed to the increased importance of Landau level mixing.

At present it is not possible to determine whether the new insulating states in the $N=1$ LL reported here bear close resemblance to either the anticipated bubble phases in the $N\ge2$ LLs or the conventional Wigner crystals in the $N=0$ level.  A distinguishing feature of the present case is that there are {\it four} re-entrant integer quantized Hall states in each spin branch of the N=1 level, two above and two below half filling.  In each spin sublevel the innermost pair of these states occur quite close to half-filling.  If these are in fact Wigner crystals, they are occuring at unusually high densities of the carriers in the valence Landau level. 

In conclusion, a number of new insulating electronic phases have been observed in the first excited Landau level of ultra-clean 2D electron systems.  The origin of the insulating behavior is quite likely a collective effect, but in the absence of theoretical predictions a detailed understanding is not yet possible. We mention in passing two intriguing preliminary findings which may assist in developing this understanding.  First, the dc current-voltage characteristics of the insulating state near $B=3.78 \ {\rm T}$ exhibit sharp and hysteretic thresholds to conduction that may reflect depinning\cite{cooper}.  Second, the insulating state near $B=5.1 \ {\rm T}$ in the lower spin branch of the $N=1$ LL exhibits small amounts of hysteresis with respect to the direction of magnetic field sweep, possibly suggesting a first order phase transition.    

This work was supported by the DOE under Grant No. DE-FG03-99ER45766 and the NSF under Grant No. DMR0070890.


\begin{references}
\bibitem{hardaxis} The ``hard" transport direction is parallel to the in-plane 
magnetic field.
\bibitem{insulate} When $\nu <1/2$ there are no Landau levels beneath the Fermi level and thus insulating electronic configurations display divergent, as opposed to vanishing, longitudinal resistance.
\end{references}
\end{document}